# Epitaxial stabilization of $Sr_3Ir_2O_7$ thin films

Junyi Yang[1,*], Lin Hao[1,*], Peyton Nanney[1], Kyle Noordhoek[1], Derek Meyers[2], Lukas Horak[3], Joshua Sanchez[4], Jiun-Haw Chu[4], Christie Nelson[5], Mark. P. M. Dean[2], Jian Liu[1,*]

1, Department of Physics and Astronomy, University of Tennessee, Knoxville, Tennessee 37996, USA;

2, Department of Condensed Matter Physics and Materials Science, Brookhaven National Laboratory, Upton, New York 11973 USA;

3, Department of Condensed Matter Physics, Charles University, Ke Karlovu 5, 121 16 Prague, Czech Republic;

4, Department of Physics, University of Washington, Seattle, Washington 98105, USA;

5, National Synchrotron Light Source II, Brookhaven National Laboratory, Upton, New York 11973, USA

*To whom all correspondence should be addressed.

J.Y.: jyang43@vols.utk.edu; L.H.: lhao3@utk.edu; J.L.: jianliu@utk.edu

**Abstract:** Ruddlesden-popper type $Sr_{n+1}Ir_nO_{3n+1}$ compound is a major focus of condensed matter physics where the subtle balance between electron-electron correlation, spin-orbit interaction and crystal field effect brings a host of emergent phenomena. While it is understandable that a canted antiferromagnetic (AFM) insulating state with an easy-plane anisotropy is developed in $Sr_2IrO_4$ as the 2D limit of the series, it is intriguing that bilayer $Sr_3Ir_2O_7$, with slightly higher effective dimensionality, stabilizes c-axis collinear antiferromagnetism. This also renders the $Sr_3Ir_2O_7$ as a unique playground to study exotic physics near a critical spin transition point. However, the epitaxial growth of the $Sr_3Ir_2O_7$ is still a challenging task because of the narrow growth window. In our research, we have studied the thermodynamic process during synthesis of $Sr_3Ir_2O_7$ thin films. We successfully expanded the synthesis window by mapping out the relation between the thin film sample crystal structure and gas pressure. Our work thus provides a more accessible avenue to stabilize metastable materials.

Layered compounds of Ruddlesden-Popper (RP) oxides $A_{n+1}B_nO_{3n+1}$ are a fertile playground to study and engineer the interplay of electronic and magnetic properties with crystal lattice dimensionality [1-5]. The crystal structure of the RP series can be viewed as $n$ consecutive $AB$O$_3$ perovskite layers sandwiched by rock-salt $A$O layers along the $c$-axis. When $n$ varies from 1 to infinity, the lattice undergoes an evolution from the quasi-two-dimensional (2D) limit to the three-dimensional (3D) limit, leading to a dimensional crossover of electronic and magnetic interactions [6-12]. For example, the 2D limit of the iridate RP family, Sr$_2$IrO$_4$ (Fig. 1(a)), represents a novel spin-orbit coupled Mott insulator [13,14]. The subtle balance between spin-orbit coupling, onsite Coulomb repulsion, and crystal field effect leads to a pseudospin $J_{\text{eff}} = 1/2$ moment on each Ir site that orders antiferromagnetically within the 2D perovskite layer with an easy $ab$-plane anisotropy and spin canting [14-17]. The magnetic structure is extremely sensitive to the dimensionality as slight increase of $n$ to 2 stabilizes a $c$-axis collinear antiferromagnetic (AFM) insulating state in Sr$_3$Ir$_2$O$_7$ (Fig.1(a)), rendering a dimensionality-controlled spin flop transition [18,19]. When further moving toward the 3D limit, the insulating ground state melts into a topological semimetallic state in paramagnetic SrIrO$_3$ (Fig.1(a)) [10-12,20-22].

To study and control the emergent phenomena within the dimensional crossover, epitaxial growth of the RP series is a particularly appealing route due to additional tunability of the lattice structure, such as imposing epitaxial strain [23-31]. However, while epitaxial thin films of Sr$_2$IrO$_4$ and SrIrO$_3$ can be readily obtained and have been characterized by many techniques [23-27,30,32-34], epitaxial growth of Sr$_3$Ir$_2$O$_7$ is much more challenging. It was reported that by using a single SrIrO$_3$ target, the Sr$_3$Ir$_2$O$_7$ phase can be obtained through carefully control of the oxygen pressure and the temperature within a small region of the parameter space [35]. The low thermodynamic stability and the narrow growth window of the Sr$_3$Ir$_2$O$_7$ phase were later confirmed by using a target of the $Sr_3Ir_2O_7$ phase [36]. On the other hand, the magnetic ordering of the $Sr_3Ir_2O_7$ thin films remains unclear.

In this work, we performed a systematic investigation of the thermodynamic stability of the iridate RP series in relation to the growth atmosphere. By using a target of the $Sr_2IrO_4$ phase, we were able to obtain high-quality thin films of single $Sr_2IrO_4$, $Sr_3Ir_2O_7$ and $SrIrO_3$ phase by varying the pure oxygen pressure. Magnetometry and magnetic resonant scattering experiments demonstrate that the obtained $Sr_3Ir_2O_7$ film preserves the same antiferromagnetic ground state as the single crystal counterpart. The obtained growth window of the $Sr_3Ir_2O_7$ phase in pure oxygen atmosphere is narrow and similar to the previous reports with $SrIrO_3$ and $Sr_3Ir_2O_7$ targets. By introducing argon into the growth atmosphere, we found a significant expansion of the growth window as a function of the oxygen partial pressure.

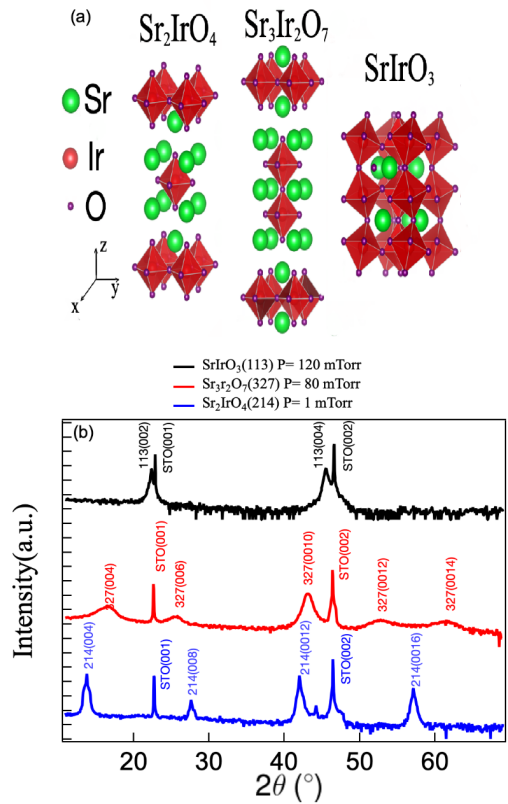

FIG. 1. (a) Schematic crystal structures of $Sr_2IrO_4$, $Sr_3Ir_2O_7$ and $SrIrO_3$. (b) X-ray Diffraction $\theta - 2\theta$ plot of representative samples with distinct RP phases.

$Sr_{n+1}Ir_nO_{3n+1}$ thin films with thickness around 100 nm were deposited on $SrTiO_3$ (001) (STO, $a_{pc}$ = 3.905 Å) single crystal substrates by using a pulsed laser deposition system (KrF excimer laser). During deposition, the substrate temperature and laser fluence were kept as 850 °C and 3 J/cm², respectively, whereas the background atmosphere composition and pressure were systematically varied. The pure oxygen atmosphere and oxygen and argon mixed gas atmosphere were chosen with gas pressure varied from 1 mTorr to 120 mTorr. Crystal structure and crystallinity of the thin films were investigated by X-ray diffraction (XRD) measurements on a Panalytical X'Pert MRD diffractometer using Cu Kα radiation (1.5406 Å). Magnetic properties measurements were carried out on a Vibrating Sample Magnetometer (Quantum Design). The dc resistivity was measured by using the standard four-point probe on a Physical Property Measurement System (Quantum Design). Synchrotron X-ray resonant magnetic scattering experiments were performed at NSLS-II beamline 4-ID, at Brookhaven National Laboratory.

Figure 1(b) shows representative XRD patterns of thin films grown in pure oxygen atmosphere but with different pressures. At the lowest pressure value used in this study (1 mTorr), only a set of (0 0 $L$) Bragg reflections that corresponds the $Sr_2IrO_4$ phase can be seen. The observation indicates epitaxial growth of $Sr_2IrO_4$ on the $SrTiO_3$ substrate along the [001] direction without observable impurity phases. The single $Sr_2IrO_4$ phase is also observed at 10 mTorr and 20 mTorr, which is consistent with previous reports [35]. At 80 mTorr and 100 mTorr, a different phase appears with all the film peaks can be indexed as the (0 0 $L$) Bragg reflections of the $Sr_3Ir_2O_7$ phase. Further increasing the growth pressure to 120 mTorr suppresses the $Sr_3Ir_2O_7$ phase and leaves only a set of film peaks near the substrate (0 0 $L$) peaks, characteristic of a single SrIrO$_3$ perovskite phase [37,38]. These results indicate that not only the $n = 1$ and $n = \infty$ structures of the RP series can be epitaxially grown by using a Sr$_2$IrO$_4$ target, the $n = 2$ structure can also be stabilized by carefully varying the atmosphere pressure.

To further elucidate on this point, we performed detailed physical properties measurements. Figure 2(a) displays the temperature dependent resistivity of the Sr$_3$Ir$_2$O$_7$ and Sr$_2$IrO$_4$ films. The monotonic increase of resistivity with decreasing temperature reveals the insulating state of both thin films. The Sr$_3$Ir$_2$O$_7$ film is evidently less insulating than the Sr$_2$IrO$_4$ film considering the smaller room-temperature resistivity and the relatively weaker insulating temperature dependence, consistent with the reports on the bulk counterparts [39-41]. The observation indicates that the dimensionality evolution of the electronic ground state of the iridate RP phases was preserved in the thin films. Figure 2(b) shows the film magnetization as a function of temperature measured under an in-plane magnetic field. The Sr$_2$IrO$_4$ film displays a weak but non-zero magnetization when temperature cools below 210 K. Note that the net magnetization of Sr$_2$IrO$_4$ is due to spin-canting of the antiferromagnetically coupled $J_{\text{eff}} = 1/2$ moments [13-16]. The magnetic measurement thus implies an antiferromagnetic transition of the Sr$_2$IrO$_4$ film with the Neel transition temperature $T_N = 210$ K. On the contrary, there is no comparable jump in magnetization in the Sr$_3$Ir$_2$O$_7$ film even down to the base temperature of 10 K.

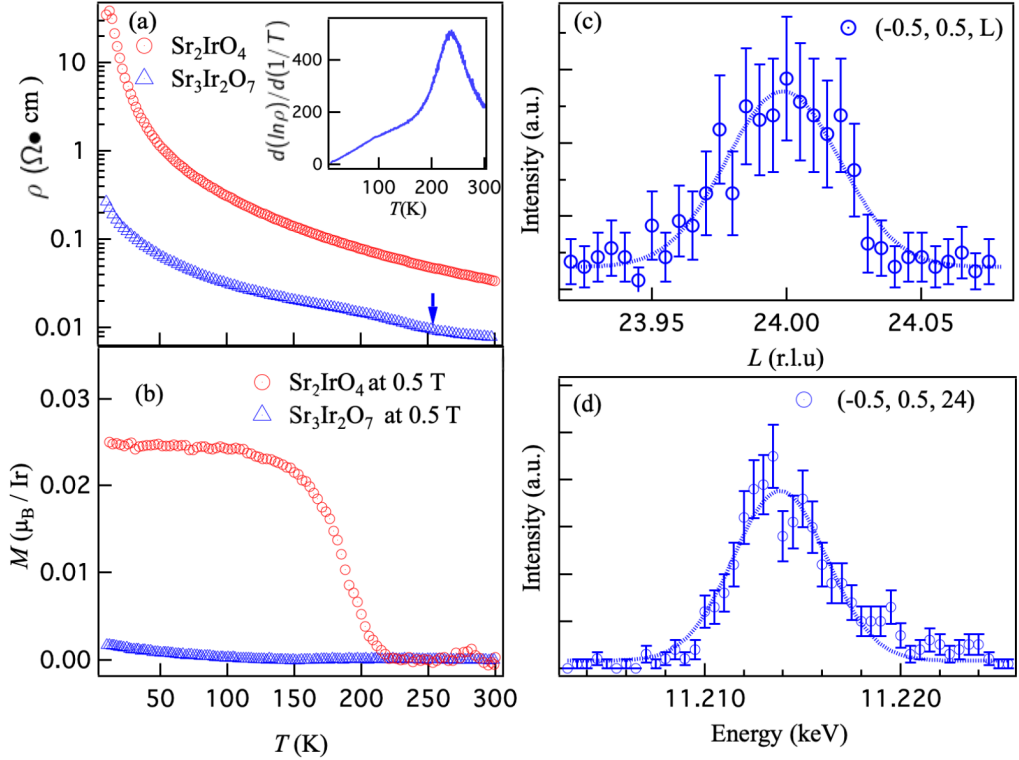

FIG. 2. (a) Temperature dependent resistivity of the $Sr_2IrO_4$ (circle) and $Sr_3Ir_2O_7$ (triangle) thin films. The resistivity kink of the $Sr_3Ir_2O_7$ thin film is indicated as a blue arrow. The temperature dependence of $d(\ln\rho)/d(1/T)$ of $Sr_3Ir_2O_7$ thin film is shown in the inset figure. A clear $\lambda$-like cup can be observed. (b) Temperature dependence of magnetic moment per Ir of $Sr_2IrO_4$ and $Sr_3Ir_2O_7$ thin films measured under a 0.5 T in-plane magnetic field. (c) $L$-scan around the (-0.5 0.5 24) magnetic reflection of the $Sr_3Ir_2O_7$ thin film at 10 K at the Ir $L_3$-edge. (d) Energy profile of the (-0.5 0.5 24) Bragg reflection peak across the Ir $L_3$-edge at 10 K. The error bars denote the statistic error.

While the absence of net magnetization along with the resistivity kink observed in the thin film are compatible with a collinear antiferromagnetic configuration as reported for $Sr_3Ir_2O_7$ single crystals [18-19,30], direct verification of the antiferromagnetic ordering is usually highly challenging for thin film samples. To this end, we exploited x-ray magnetic resonant scattering, which has been proven to be a powerful probe of magnetic structure of iridates due to the element selectivity and resonant enhancement [13]. By tuning the x-ray photon energy to the Ir $L_3$-edge, we performed magnetic resonant scattering measurements on the $Sr_3Ir_2O_7$ film at 10 K. The (-0.5 0.5 24) AFM Bragg peak was clearly identified by the $L$-scan, as shown in Fig. 2(c), directly demonstrating that the Ir moments are antiferromagnetically ordered within the $IrO_2$ plane. Fig. 2(d) shows the energy profile at the representative magnetic reflection across the Ir $L_3$-edge. A clear resonant effect can be seen, at energy slightly lower than the Ir $L_3$ white line similar to other iridium compounds [13,42-43], confirming the dominant role of Ir ions in developing the long-range magnetic ordering. This observation agrees well with the G-type AFM structure as reported in a $Sr_3Ir_2O_7$ single crystal [18-19, 44-45]. Along with the structural analysis, we conclude that the as-obtained film under the intermediate pressure indeed has a single $Sr_3Ir_2O_7$ phase.

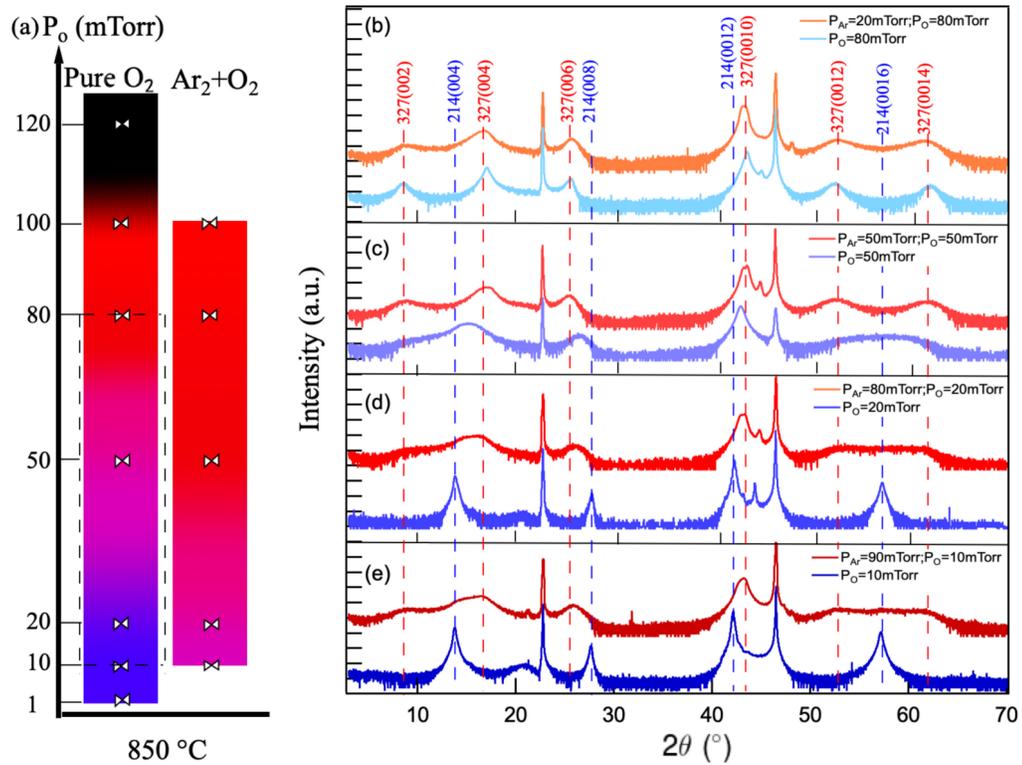

FIG. 3. (a) Growth phase diagram of $Sr_{n+1}Ir_nO_{3n+1}$ thin films. The left panel summaries the results under pure oxygen atmosphere and the dashed rectangle highlights the interested oxygen pressure range from 1 to 120 mTorr. The right panel shows results under mixed oxygen and argon atmosphere. The black region presents 113 phases, red region stands for 327 phases and blue region stands for 214 phases. Batwing markers label synthesized thin films. X-ray Diffraction $\theta - 2\theta$ plot for thin film samples synthesized under $P_O = 80$ mTorr (b), 50 mTorr (c), 20 mTorr (d) and 10 mTorr (e) in both mixed argon and pure oxygen gas atmosphere.

The growth evolution of the RP phases as a function of oxygen atmosphere pressure is summarized in the left panel of Fig. 3(a). As the oxygen pressure increases from 1 mTorr up to 100 mTorr, the obtained thin film evolves from a single $Sr_2IrO_4$ phase to a single $Sr_3Ir_2O_7$ phase. Between these two single phases, there is an intermediate region where a mixed phase is obtained. The growth window of the pure $S_3Ir_2O_7$ phase is relatively narrow and spans a range of ~ 20 mTorr only. To enlarge the growth window, we systematically tuned the oxygen partial pressure while fixing the total atmosphere pressure to be 100 mTorr by introducing different amount of argon into the chamber. Five oxygen partial pressure, 80 mTorr, 50 mTorr, 20mTorr, 10 mTorr and 0 mTorr, were selected for this study. Starting from high oxygen partial pressure $P_o = 80$ mTorr ($P_{Ar}=20$ mTorr), the obtained thin film displays a single $Sr_3Ir_2O_7$ phase (Fig. 3 (b)), which is the same as that in pure oxygen atmosphere with $P_o = 80$ mTorr. By decrease the $P_O$ down to 50 mTorr ($P_{Ar} = 50$ mTorr), interestingly, the obtained thin film still only shows the set of Bragg reflections that characterizes a $Sr_3Ir_2O_7$ single phase. This is in sharp contrast with the mixed phase film synthesized under pure oxygen atmosphere of the same $P_o = 50$ mTorr (Fig. 3 (c)). As the $P_O$ further decreases down to 20 mTorr ($P_{Ar} = 80$ mTorr) and 10 mTorr ($P_{Ar} = 90$ mTorr), the $Sr_3Ir_2O_7$ phase remains robust though the Bragg reflections are broadened (Fig. 3 (c) and (d)). In other words, the thermal stability of the $Sr_3Ir_2O_7$ phase has been significantly enhanced by introducing argon. On the other hand, in a pure argon atmosphere, i.e., $P_o = 0$ mTorr, none of the above RP phases can be synthesized, highlights the critical role of oxygen in stabilizing a $Sr_{n+1}Ir_nO_{3n+1}$ phase. The growth evolution as a function of $P_o$ within mixed atmosphere is sketched in the right panel of Fig. 3(a). As compared to that obtained in pure oxygen atmosphere, it is clear that the growth window of the $Sr_3Ir_2O_7$ phase has been greatly expanded after introduction of argon.

From the chemical formula $A_{n+1}B_nO_{3n+1}$ of the RP oxides, it can be seen that the $B/A$ cation ratio increases from *0.5* for $n = 1$ to *1* for $n = \infty$. The $Sr_2IrO_4$ and $Sr_3Ir_2O_7$ phases can be considered as variants of $SrIrO_3$ with different degrees of Ir-deficiency. It is indeed possible for $SrIrO_3$ to decompose into the various RP members with the byproduct of Ir and $O_2$, or vice versa. As shown by the previous studies [35,36], such a controllability of the thermodynamic stability of the three RP phases can be achieved during the growth by varying the ambient pressure of pure oxygen gas, which is also observed in our study. On the other hand, the background gas pressure of pulsed laser deposition is also known to strongly influence the plasma plume dynamics, including the ratio and energetics of different ions [46-47]. This effect may also have significant impact to the growth kinetics, such as the sticking coefficients of different ions and species, especially when the pressure is tuned by more than two orders magnitudes [46]. Such impact is confirmed by the observed expansion of the growth window of the $Sr_3Ir_2O_7$ phase when the overall pressure is maintained by introducing the Ar gas. In other words, when reducing the pressure under a pure oxygen atmosphere, the changes in the thermodynamic phase stability and the plume dynamic both favor the conversion from the Ir-rich to Ir-deficient phases. This combination results in a sharp evolution between the two end members, the $Sr_2IrO_4$ and $SrIrO_3$ phases with a narrow window of the $Sr_3Ir_2O_7$ phase in between. Indeed, the previous and our studies all found a remarkably similar phase dependence on the oxygen pressure from the level of 1 mTorr to 100 mTorr regardless of the target stoichiometry. Such a phase evolution is significantly slowed down when the plume dynamics is stabilized by the addition of the Ar partial pressure, extending the growth window of the $Sr_3Ir_2O_7$ phase.

In conclusion, we systematically investigated the effect of growth atmosphere on the epitaxial growth of $Sr_{n+1}Ir_nO_{3n+1}$ series. The magnetic scattering measurements in combining with structural analysis and physical properties measurement enable us to draw the growth phase diagram as a function of oxygen pressure, upon which the narrow growth window of the $Sr_3Ir_2O_7$ phase is highlighted. We demonstrated that this growth window can be greatly expanded through introducing argon in the growth chamber. Known that pure oxygen is widely used during oxide synthesis process, the present study affords an efficient route to synthesize a metastable phase during epitaxial growth.

The authors acknowledge experimental assistance from H.D. Zhou, M. Koehler, J.K. Keum. J.L. acknowledges support from the Science Alliance Joint Directed Research & Development Program and the Organized Research Unit Program at the University of Tennessee. M.P.M.D is supported by the U.S. Department of Energy, Office of Basic Energy Sciences, Early Career Award Program under Award No. 1047478. J.S. and J.-H.C. are supported by the Air Force Office of Scientific Research Young Investigator Program under Grant FA9550-17-1-0217. Work at Brookhaven National Laboratory was supported by the U.S. Department of Energy, Office of Science, Office of Basic Energy Sciences, under Contract No. DESC00112704. This research used resources at the 4-ID-1 beam line of the National Synchrotron Light Source II, a U.S. Department of Energy (DOE) Office of Science User Facility operated for the DOE Office of Science by Brookhaven National Laboratory under Contract No. DE-SC0012704.

**References**:

[1] J.G. Rau, E.K.H. Lee and H.Y. Kee, Annu. Rev. Condens. Matter Phys. **7**, 195 (2015).

[2] G. Cao and P. Schlottmann, Rep. Prog. Phys. **81** 042502 (2018).

[3] S. J. Moon, H. Jin, K. W. Kim, W. S. Choi, Y. S. Lee, J. Yu, G. Cao, A. Sumi, H. Funakubo, C. Bernhard, and T. W. Noh, Phys. Rev. Lett. **101**, 226402 (2008).

[4] Q. Wang, Y. Cao, J.A. Waugh, S.R. Park, T.F. Qi, O.B. Korneta, G. Cao, and D.S. Dessau, Phys. Rev. B **87**, 245109 (2013).

[5] I. Pallecchi, M. T. Buscaglia, V. Buscaglia, E. Gilioli, G. Lamura, F. Telesio, M. R. Cimberle, and D. Marre, J. Phys.: Condens. Matter **28**, 065601 (2016).

[6] G. Cao, J. Bolivar, S. McCall, J.E. Crow and R.P. Guertin, Phys. Rev. B **57**, R11039(R) (1998).

[7] S. Chikara, O. Korneta, W. P. Crummett, L. E. DeLong, P. Schlottmann, and G. Cao, Phys. Rev. B **80**, 140407(R) (2009).

[8] S. Fujiyama, H. Ohsumi, T. Komesu, J. Matsuno, B. J. Kim, M. Takata, T. Arima, and H. Takagi, Phys. Rev. Lett. **108**, 247212 (2012).

[9] S. J. Moon, H. Jin, W. S. Choi, J. S. Lee, S. S. A. Seo, J. Yu, G. Cao, T. W. Noh, and Y. S. Lee, Phys. Rev. B **80**, 195110 (2009).

[10] G. Cao, V. Durairaj, S. Chikara, L. E. DeLong, S. Parkin, and P. Schlottmann, Phys. Rev. B **76**, 100402(R) (2007).

[11] Y.F. Nie, P.D.C. King, C.H. Kim, M. Uchida, H.I. Wei, B.D. Faeth, J.P. Ruf, L. Xie, X. Pan, C.J. Fennie, D.G. Schlom and K.M. Shen, Phys. Rev. Lett. **114**, 016401 (2015).

[12] J. Fujioka, T. Okawa, A. Yamamoto, and Y. Tokura, Phys. Rev. B **95**, 121102(R) (2017).

[13] B.J. Kim, H. Ohsumi, T. Komesu, S. Sakai, T. Morita, H. Takagi and T. Arima, Science **323** 1329 (2009).

[14] B.J. Kim, H. Jin, S.J. Moon, J-Y. Kim, B.G. Park, C.S. Leem, J. Yu, T.W. Noh, C. Kim, S.J. Oh, J.H. Park, V. Durairaj, G. Cao, and E. Rotenberg, Phys. Rev. Lett. **101**, 076402 (2008).

[15] G. Jackeli and G. Khaliullin, Phys. Rev. Lett. **102**, 017205 (2009).

[16] F. Wang and T. Senthil, Phys. Rev. Lett. **106**, 136402 (2011).

[17] H. Jin, H. Jeong, T. Ozaki, and J. Yu, Phys. Rev. B **80**, 075112 (2009).

[18] J. W. Kim, Y. Choi, Jungho Kim, J. F. Mitchell, G. Jackeli, M. Daghofer, J. van den Brink, G. Khaliullin, and B. J. Kim, Phys. Rev. Lett. **109**, 037204 (2012).

[19] S. Boseggia, R. Springell, H.C. Walker, A.T. Boothroyd, D. Prabhakaran, D. Wemeille, S.P. Collins and D.F. McMorrow, J. Phys.: Condens. Matter **24** 312202 (2012).

[20] J.M. Carter, V.V. Shankar, M. Ahsan Zeb and H.Y. Kee, Phys. Rev. B **85**, 115105 (2012).

[21] Z. T. Liu, M. Y. Li, Q. F. Li, J. S. Liu, W. Li, H. F. Yang, Q. Yao, C. C. Fan, X. G. Wan, Z. Wang and D. W. Shen, *Sci. Rep.* **6**, 30309 (2016).

[22] M. Ahsan Zeb and Hae-Young Kee, Phys. Rev. B **86**, 085149 (2014).

[23] A. Biswas, K.-S. Kim and Y. H. Jeong, J. Appl. Phys. **116**, 213704 (2014).

[24] K. R. Kleindienst, K. Wolff, J. Schubert, R. Schneider and D. Fuchs, Phys. Rev. B **98**, 115113 (2018).

[25] Jian Liu, D. Kriegner, L. Horak, D. Puggioni, C. Rayan Serrao, R. Chen, D. Yi, C. Frontera, V. Holy, A. Vishwanath, J. M. Rondinelli, X. Marti, and R. Ramesh, Phys. Rev. B **93**, 085118 (2016).

[26] C. Rayan Serrao, Jian Liu, J. T. Heron, G. Singh-Bhalla, A. Yadav, S. J. Suresha, R. J. Paull, D. Yi, J.-H. Chu, M. Trassin, A. Vishwanath, E. Arenholz, C. Frontera, J. Železný, T. Jungwirth, X. Marti, and R. Ramesh, Phys. Rev. B **87**, 085121 (2013).

[27] J. Nichols, O. B. Korneta, J. Terzic, L. E. De Long, G. Cao, J. W. Brill, and S. S. A. Seo, Appl. Phys. Lett. **103**, 131910 (2013).

[28] B. Kim, P. Liu, and C. Franchini, Phys. Rev. B **95**, 024406 (2017).

[29] A. Lupascu, J. P. Clancy, H. Gretarsson, Z. Nie, J. Nichols, J. Terzic, G. Cao, S. S. A. Seo, Z. Islam, M. H. Upton, J. Kim, D. Casa, T. Gog, A.H. Said, V.M. Katukuri, H. Stoll, L. Hozoi, J. van den Brink, and Y-J. Kim, Phys. Rev. Lett. **112**, 147201 (2014).

[30] J. Nichols, J. Terzic, E. G. Bittle, O. B. Korneta, L. E. D. Long, J. W. Brill, G. Cao, and S. S. A. Seo, Appl. Phys. Lett. **102**, 141908 (2013).

[31] L. Miao, H. Xu, and Z. Q. Mao, Phys. Rev. B **89**, 035109 (2014).

[32] C. L. Lu, A. Quindeau, H. Deniz, D. Preziosi, D. Hesse, and M. Alexe, Appl. Phys. Lett. **105**, 082407 (2014).

[33] O. Krupin, G. L. Dakovski, B. J. Kim, J. W. Kim, J. Kim, S. Mishra, Y.-D. Chuang, C. R. Serrao, W. S. Lee, W. F. Schlotter, M. P. Minitti, D. Zhu, D. Fritz, M. Chollet, R. Ramesh, S. L. Molodtsov, and J. J. Turner, J. Phys.: Condens. Matter **28**, 32LT01 (2016).

[34] M. P. M. Dean, Y. Cao, X. Liu, S. Wall, D. Zhu, R. Mankowsky, V. Thampy, X. M. Chen, J. G. Vale, D. Casa, Jungho Kim, A. H. Said, P. Juhas, R. Alonso-Mori, J. M. Glownia, A. Robert, J. Robinson, M. Sikorski, S. Song, M. Kozina, H. Lemke, L. Patthey, S. Owada, T. Katayama, M. Yabashi, Yoshikazu Tanaka, T. Togashi, J. Liu, C. Rayan Serrao, B. J. Kim, L. Huber, C.-L. Chang, D. F. McMorrow, M. Först & J. P. Hill, *Nat. Mater.* **15**, 601–605 (2016).

[35] K. Nishio, H.Y. Hwang, and Y. Hikita, APL Mater. **4**, 036102 (2016).

[36] A. Gutiérrez-Llorente, L. Iglesias, B. Rodríguez-González, and F. Rivadulla, APL Mater. **6**, 091101 (2018).

[37] J.M. Longo, J.A. Kafalas, R,J, Arnott, J. Solid State Chem. **3**, 174, (1971).

[38] L. Horák, D. Kriegner, J. Liu, C. Frontera, X. Martib and V. Holý, J. Appl. Cryst. **50**, 385 (2017).

[39] G. Cao, Y. Xin, C.S. Alexander, J.E. Crow, P. Schlottman, M.K. Crawford, R.L. Harlow and W. Marshall, Phys. Rev. B **66**, 214412 (2002).

[40] S. Fujiyama, K. Ohashi, H. Ohsumi, K. Sugimoto, T. Takayama, T. Komesu, M. Takata, T. Arima, and H. Takagi, Phys. Rev. B **86**, 174414 (2012).

[41] N.S. Kini, A.M. Strydom, H.S. Jeevan, C. Geibel and S. Ramakrishnan, J. Phys.: Condens. Matter **18** 8205 (2006).

[42] S. Boseggia, R. Springell, H. C. Walker, H. M. Rønnow, Ch. Rüegg, H. Okabe, M. Isobe, R. S. Perry, S. P. Collins, and D. F. McMorrow, Phys. Rev. Lett. **110**, 117207 (2013).

[43] H. Sagayama, D. Uematsu, T. Arima, K. Sugimoto, J. J. Ishikawa, E. O'Farrell, and S. Nakatsuji, Phys. Rev. B **87**, 100403(R) (2013).

[44] S. Boseggia, R. Springell, H. C. Walker, A. T. Boothroyd, D. Prabhakaran, D. Wermeille, L. Bouchenoire, S. P. Collins, and D. F. McMorrow, Phys. Rev. B **85**, 184432 (2012).

[45] C. Dhital, S. Khadka, Z. Yamani, C. de la Cruz, T. C. Hogan, S. M. Disseler, M. Pokharel, K. C. Lukas, W. Tian, C. P. Opeil, Z. Wang, and S. Wilson, Phys. Rev. B **86**, 100401(R) (2012).

[46] R. Eason, *Pulsed Laser Deposition of Thin Films: Applications-Led Growth of Function Materials*, John Wiley & Sons, Inc. HoboKen, New Jersey, 2007

[47] J.C. Miller and R. F. Haglund Jr., *Experimental Methods in the Physical Sciences, Volume 30, Laser Ablation and Deposition,* Academic Press, San Diego, 1998